# A New Approach to Detect Important Members that Create the Communities in Bipartite Networks

Ali Hojjat
Department of Industrial Engineering, Tarbiat Modares University, Tehran, Iran
alihojjat@modares.ac.ir

Ghazaleh Haddad
Department of Industrial Engineering, Tarbiat Modares University, Tehran, Iran
ghazaleh.haddad@modares.ac.ir

*Abstract*—The world around us consists of objects that have different relationships with each other. The result of these communications is various networks, part of which are bipartite networks. While many studies have investigated essential network members, less attention has been paid to the bipartite graphs. On the other hand, one of the most critical aspects of network analysis is the detection and extraction of communities that arise in the structure of networks. For these reasons, we have introduced a measure called H.H to identify influential nodes in community formation in the one-mode projection of a bipartite graph. The three main parameters that influence this measure are the size of the formed community, the effect of each node in the formation of that community, and the number of communities in which the node had an impact. The results of this paper show the differences of this measure with other centralities (eigenvector centrality, closeness centrality, betweenness centrality, and degree centrality) and how this measure takes into account aspects that other centralities do not. Through H.H score, we can find essential nodes that have been effective in forming a community, and by removing these nodes, communities can be eliminated. On the other hand, by adding nodes with a good H.H score, more important and more vital communities can be created in the one-mode projection of a bipartite graph. Any of the existing centralities has not addressed this issue, and this measure has sufficient independence to represent the important nodes in the formation of the communities. Experimental validation of the proposed measure is carried out on two real-world datasets: Southern Women Network and Person-Crime Network. The results of the implementation of the H.H score on the Person-Crime dataset show that by eliminating the nodes with the highest H.H score (top-10%), 29% of the communities have changed; this is while the centralities change the average of 18% of the communities and this shows the importance of the H.H score.

## I. INTRODUCTION AND BACKGROUND

Objects and the environment around them can be expressed as a network based on their relationships and some members are more important because of their situation in the network. For example, a member who knows more people can more quickly send a message to all network members. In the networks, there are two basic concepts, called nodes and edges. The nodes are the actors, and the edges are communications between them [1]. The connection of a node to other nodes in a network can specify information about the importance or unimportance of that node in specific applications. To measure this importance, we use quantitative indicators that are called centrality measures. Over the years, various types of centrality measures have been introduced in a graph, which determines the relative importance in a graph [2]. Currently, there are several measures for calculating centralities. The concept of centrality refers to the position of a node in a network, which is understood in the structure of the network [3]. Therefore, it can be said that each node that has a better position in the network gets a higher score from centrality measures. Freeman has introduced three types of centrality that are widely used in network analysis: degree centrality, closeness centrality, and betweenness centrality. Each of them examines the position of a node in the network structure from a different perspective [4].

The world around us consists of a variety of networks: social, political, environmental, ecological, and so on, which have different members and are linked for various reasons, but many of these networks are two types or gender of nodes. For example, actors and events networks are defined as such; if someone participates in an event, they will communicate between the two types of nodes and create the network. These types of networks are called bipartite networks. That is, if we can divide nodes of a graph into two subsets, which are all communications between the two sets, we define this graph as bipartite [5][6][7][8]. As shown in (Fig.1), nodes of this type of graph can be defined as a set of individuals "A" and a set of events "E" which set |E+A| is the total number of members or nodes of the graph.

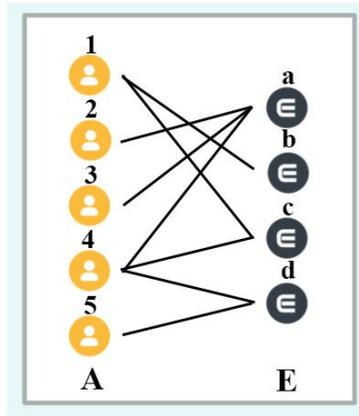

Fig.1. example of a bipartite graph

As mentioned, there are many centrality measures to examine the position of a node in the network. These centralities are the most used measures in studies of network analysis [4]. On the other hand, if the nodes of the network are easily divided into groups where the nodes of each group are closely interconnected, and the groups themselves have little connection, the network is made of various communities. One of the most important aspects of analyzing networks is the recognition and extraction of these communities from the structure of these networks. The effective node in forming a community may have a significant and functional role in controlling and sustaining community members [9].

Therefore, community detection is also considered an essential networking tool in communication analysis. However, can this powerful tool be used to identify crucial nodes in the network?

This study aims to identify critical or influential members that create the communities in bipartite networks using community detection. Commonly the community structure of only one of the modes is analyzed, and the bipartite network is projected into a one-mode network (one-mode projection) [10]. For this reason, we convert the bipartite graph to an item-item graph (one-mode projection) and identify the communities in this graph. The reason for this action is that an item-item graph can effectively extract hidden relationships between nodes within the same vertex sets of bipartite graphs [11]. In the next step, due to the effect of each vertex in forming a community, we give a score to each node. We called this score H.H. We will use two real-world bipartite network datasets, person-crime and Southern Women, to evaluate this measure.

## II. METHODS

We used the Southern Women Network dataset [12] (fig.2) to implement the H.H measure. In this figure, the size of the nodes is based on their degree in the network. This dataset is a network that shows the connection between 18 women (actors) and 14 social events. Nodes of this network can be divided into two sets of women and events, and only the edges between the two sets are created and make a bipartite network. This network has 89 edges and 32 nodes. This data is usually used to identify communities in bipartite graphs [13].

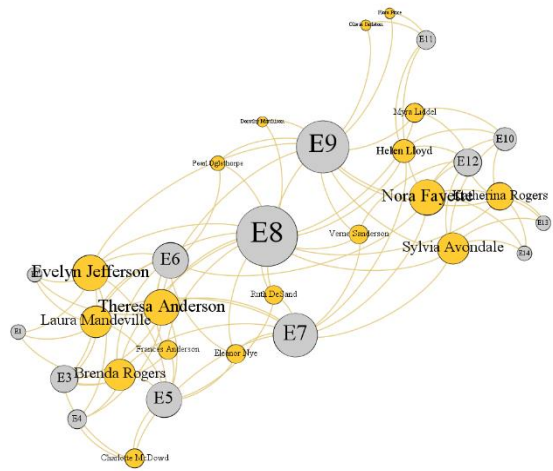

Fig.2. Southern Women Network

**1. One-Mode Projections of Bipartite graph:**

Communities in a bipartite graph may be viewed from different perspectives [14][15][16]. Commonly the community structure of only one of the modes is analyzed [11][10]. For this reason, to discover communities in bipartite graphs, we first convert the bipartite graph (actor-event) into an item-item graph (actor-actor or event-event). For example, if we have the edge (event a, actor 2) and the edge (event a, actor 3), then in the actor-actor graph, we will have the edge (actor 2, actor 3) (fig.3.2). But it should be noted that in the one-mode projection, part of the data will be lost [17].

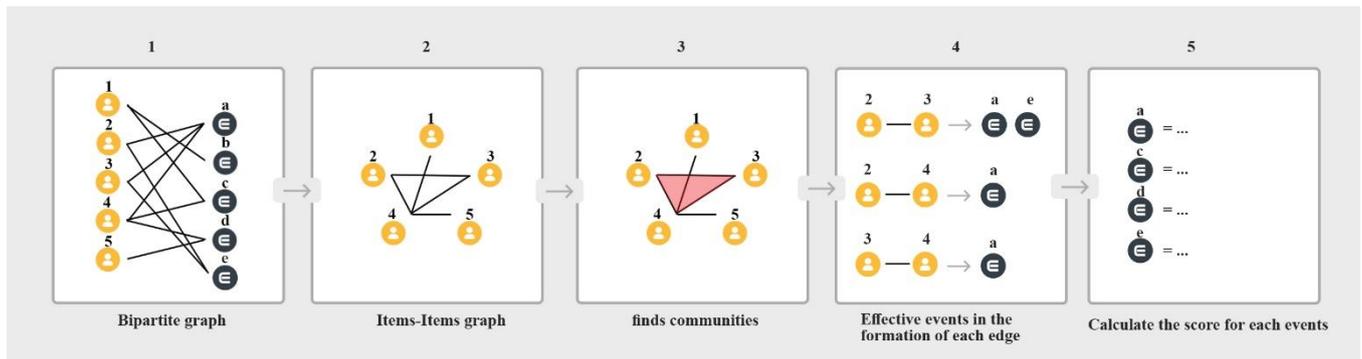

Fig.3. Step 1: Bipartite Graph, Step 2: Bipartite Graph Conversion to actor-actor Graph, Step 3: Finding Communities, Step 4: Effective events in the formation of each edge, Step 5: Calculate H.H Score for Each event

**2. Extract communities in One-Mode Projections of Bipartite graph:**

After creation a one-mode projection of a bipartite graph, we find communities in it. We use the Maximum clique [18] to define these communities. Nodes may be common among cliques or communities (fig.3.3).

Now, our goal is to understand which nodes are more effective in creating a community. The reason for the formation of a community in the actor-actor graph is participation in common events. Because of that, we extract the neighbors of each node from the main bipartite graph (the events that each actor participates in). The more an event is effective in creating more edges, the greater its impact on community formation.

**3. Introducing H.H measure:**

After extracting communities, our goal is to introduce a measure (H.H) that shows the effect of each event in forming a community in the one-mode projection of bipartite graphs (fig.3).

To define the measure for each node, we must answer three questions:
1. How many communities in one-mode projection (actor-actor graph) have been affected by each event (node)?
2. What is the size of the community (in actor-actor graph) that has been created?
3. What is the effect of each event (node) in the formation of each community (in actor-actor graph)?

Accordingly, the three main parameters of the H.H are as follows:
1. *How many communities have been affected by each event?*

    As said, nodes may be common among communities; therefore, these commons should be considered in the calculation of the H.H score. For this reason, we calculate scores for each node (event) in all communities that include it, and finally, we sum these scores. Thus, with this approach, we consider the effect of each "event" in all communities that contain it.

    For example, in (fig.4), event "f" has been influential in forming two communities (green and red), so the scores obtained for this node in the green and red communities are separately calculated, and finally, we sum these scores.

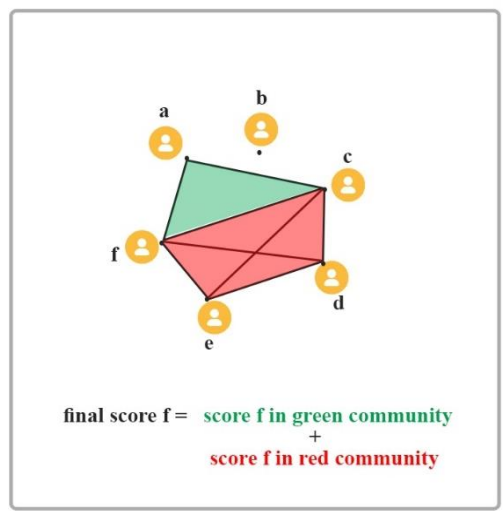

Fig.4 The scores earned by each node are summed for all communities that including it

$$Score\ a = score\ a\ in\ green\ community$$
$$Score\ b = 0$$
$$Score\ c = score\ c\ in\ green\ community + score\ c\ in\ red\ community$$
$$Score\ d = score\ d\ in\ red\ community$$
$$Score\ e = score\ e\ in\ red\ community$$
$$Score\ f = score\ f\ in\ green\ community + score\ f\ in\ red\ community$$

2. *What is the size of the community that has been created?*
   As shown in Eq.1, the magnitude of the number of edges (connections) in a community with n nodes is $O(n^2)$. This means in the case of the disappearance of a nodes (event) that affect in creation of a larger community (in actor-actor graph) more edges will be vanished. Accordingly, as the community's size (in actor-actor graph) grows, the score of the nodes (events) that created that community should be increased.

$$Number\ of\ community\ (clique)\ edges \cong \frac{n*(n-1)}{2} = O(n^2) \quad (1)$$

   For example, in a community with 3 members, if the node that made the community disappears, we have lost three edges, but if the size of the community that created is 30 (10 times larger) and the node that made the community goes away, we have lost 435 (145 times larger) edges. This indicates the importance of the size of the community. To show the number of members of the formed community in Eq.3, we used $k_j$.

3. *What is the effect of each event in the formation of each community (in actor-actor graph)?*
   To understand this, we first need to look at why a community is created in the first place. As showed in (fig.5), the reason for the formation of the community in the actor-actor graph **is the presence of common events among the community members**. To calculate the impact of each event, we can calculate the impact of each event on the formation of each community edge; if an event creates an edge that no other event has done, then it gets a higher score, but if an event creates an edge that many other events make, the score will be shared between all of them. For example, in Figure 5, the community {1, 2, 3} has been created. In this community, we have 3 edges: {1-2 ,1-3 ,2-3}. Event 'a' creates edges '1-2' and '1-3', and no other event is effective in forming these edges. For this reason, removing the event 'a' eliminates these edges and may affect the community. So event 'a' gets the full score because it is the only reason to create these edges. But on the other hand, in creating the edge '2-3', 'a' and 'c' are effective and if one of these events disappears, the other event will create the edge '2-3'. That's why the scores are shared between two events 'a' and 'c'. In other words, if different events create an edge of the community, the score is shared between these events.

   To find this effect, we can use this ratio:

$$effect\ of\ event_i\ in\ the\ formation\ of\ Community_j = \sum_{edge \in Community_j} \frac{1}{The\ number\ of\ events\ that\ make\ that\ edge} \quad (2)$$

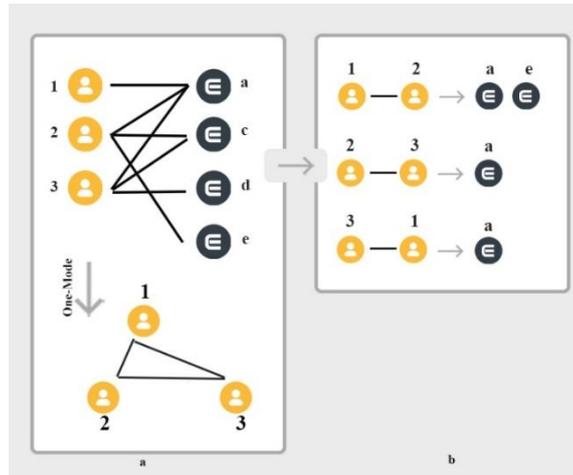

Fig.5-a: The events that each actor participates in   5-b: Effective events in the formation of each year

Suppose each actor goes to different events:

$$actor_i \in \{event_{x1}, event_{x2}, event_{x3} \dots, event_{xk}\}$$

The score for each event is calculated as follows:

$$H.H\ (event_i) = \sum_{c_j \in C} k_j * \sum_{edge \in c_j} \frac{1}{w} \quad (3)$$

$C = set\ of\ graph\ communities$
$c_j = Community_j$
$k_j = Size\ of\ Community_j$
$w = The\ number\ of\ events\ that\ make\ that\ edge$

As we said, there are various measures for identifying essential nodes in the graphs, in this section, we compare these measures with the H.H score:

**Degree centrality**

Using the adjacency matrix $A = (a_{ij})$ Degree centrality can be formalized as follows [19]:

$$\sigma_d\ a_{ij} = \sum_{i=1}^{n} a_{ix} \quad (4)$$

According to Eq(3), Degree centrality is associated with the node's degree. On the other hand, because the node's degree effectively creates a community in a one-mode projection, if there is a node with a high degree of centrality in the bipartite graph, then that node can have a large H.H score. But this is not always the case, and these two measures are different from each other. To understand this difference, we explain an example. The difference between the graphs in Fig. 6-a and Fig. 6-b are red edges. In these two graphs, the degree centrality of event '2' has not changed, but its H.H score has changed. The reason for this is that some of the edges that event '2' has created in the community (like edge 'bc'), are now being created by other events (events '3' and '4'). For this reason, the impact of '2' on the formation of community edges has decreased, and thus its score fell by 26%. On the other hand, since node '3' and node '4' have been effective in forming one of the edges of the community (edge 'bc'), their H.H score has been increased.

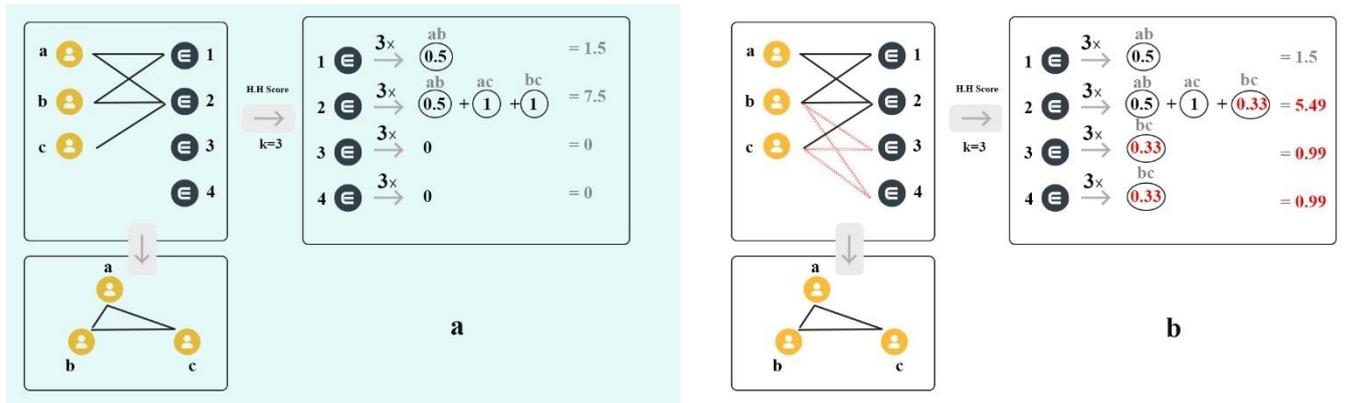

Fig.6-a: Calculate the H.H score for each node  6-b: The addition of 4 edges to the graph causes the score of some of the events to be changed while their degree is constant.

**betweenness centrality**

with $g_{ij}(x)$ representing the number of shortest paths from node $i$ to node $j$, and $g_{ij}(x)$ denoting the number of these paths which pass through the node $x$, degree centrality can be formalized as follows [20]:

$$\sigma_b(x) = \sum_{i=1, i \neq x}^{1} \sum_{j=1, j<i, j \neq x}^{1} \frac{g_{ij}(x)}{g_{ij}} \quad (\circ)$$

This centrality is related to the number of times a node is located in the shortest path between the two other nodes. But the H.H score is related to the effect of a node on community formation in the one-mode projection. Therefore, there is no connection between the H.H score and betweenness centrality. It should be noted that a node with a very good position (based on betweenness centrality) may have a good H.H score. Because if the betweenness centrality of a node is good, then that node is generally in a good position and can also have good scores from the perspective of other centralities (closeness, eigenvector, degree) and H.H score.

**Closeness centrality**

Closeness centrality checks nodes in terms of their distance from other nodes [3]. On the other hand, the H.H score examines the impact of a node on community formation in the one-mode projection graph. For this reason, these two measures have no relation with each other. As mentioned for betweenness centrality, a node with a very good position (based on closeness centrality) may have a good H.H score.

$$\sigma_c(x) = \frac{1}{\sum_n^1 distance_G(x.i)} \tag{6}$$

**Eigenvector centrality**

The eigenvector centrality of each vertex is proportional to the sum of the centralities of its neighbors [21], So the node's degree is effective in eigenvector centrality. Also, the node's degree is effective in the H.H score. Therefore, we can say that there is little correlation between the eigenvector centrality and H.H score, but its strength is not as strong as the correlation between degree centrality and H.H score.

## III. RESULT AND DISCUSSION

We were able to present a new measure to identify the important members of the bipartite network. This measure is defined using the effect of each node in the formation of communities in the one-mode projection of bipartite graphs. In this section, we will evaluate the H.H score in two datasets.

To compare H.H score with other centralities; first, we normalize the numbers of each centrality with Eq. (7):

$$nx_i = \frac{x_i - min}{max - min} \tag{7}$$

### 1. Southern Women Dataset

In order to implement the H.H score, we used real-world data of southern women. This dataset is a network that shows the connection between 18 women and 14 social events. We first convert the bipartite graph to an actor-actor graph (one-mode projection). Next, we find communities in the actor-actor graph. (Fig.7)

After recognizing the communities in the actor-actor graph, we evaluate the events according to the H.H score. (Table.1)

Table.1. H.H score of events

|     | Normalized Score | Score     |
| --- | ---------------- | --------- |
| E1  | 0                | 13.5      |
| E2  | 0.0010           | 14.7142   |
| E3  | 0.0532           | 75.2142   |
| E4  | 0.0096           | 24.7142   |
| E5  | 0.1515           | 189.2976  |
| E6  | 0.2099           | 256.9642  |
| E8  | 1                | 1173.1309 |
| E9  | 0.7111           | 838.1476  |
| E7  | 0.4132           | 492.75    |
| E12 | 0.1041           | 134.2666  |
| E10 | 0.0564           | 78.9333   |
| E13 | 0.0029           | 16.9333   |
| E14 | 0.0029           | 16.9333   |
| E11 | 0.0362           | 55.5      |

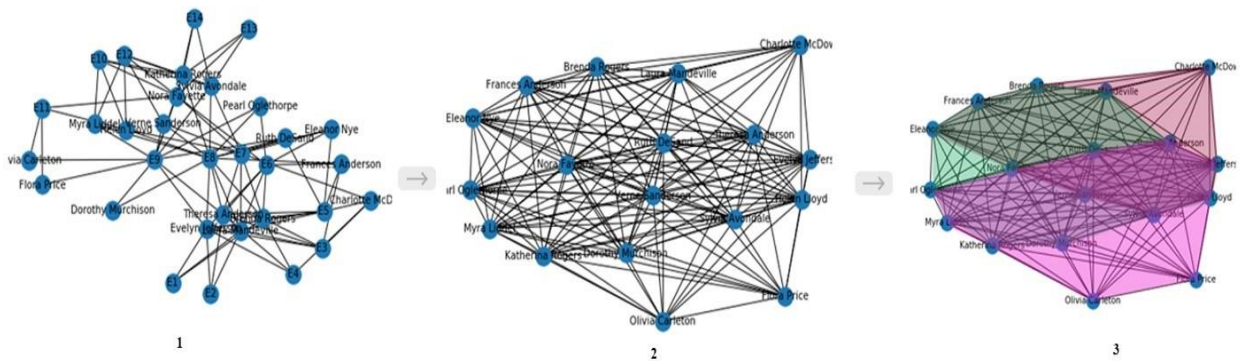

Fig.7. Step 1: Bipartite Graph, Step 2: Bipartite Graph Convert to actor-actor Graph, Step 3: Finding Communities

To evaluate the performance of the H.H measure, we compare it with the degree centrality, closeness centrality, betweenness centrality, and eigenvector centrality. Communities created in this graph represents the actors who have been together in a large number of events (Fig.7-3)

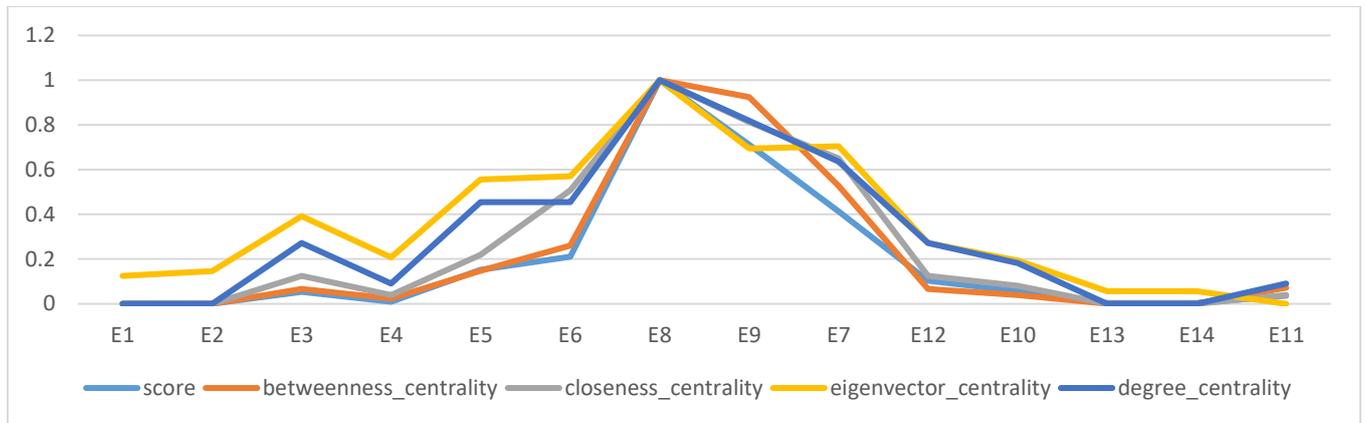

Fig.8. Comparison of H.H with other centralities (normalized)

As shown in (fig.8), changes in the H.H score are almost similar with centralities. But this is not always the case. For example, the degree centrality of 'E6' is equal to 'E5', but 'E6' has a higher H.H score because in calculating the H.H score, three parameters are effective, and these parameters are the main reason for the difference. In other words, the 'E6' has been effective in forming more essential edges in communities.

## 2. Person-Crime Dataset

This bipartite network contains persons who appeared in at least one criminal case as either a suspect, a victim, a witness, or both a suspect and victim at the same time. An edge between two nodes shows that the person was involved in the crime (551 persons, 829 crimes, 1478 edges) [22].

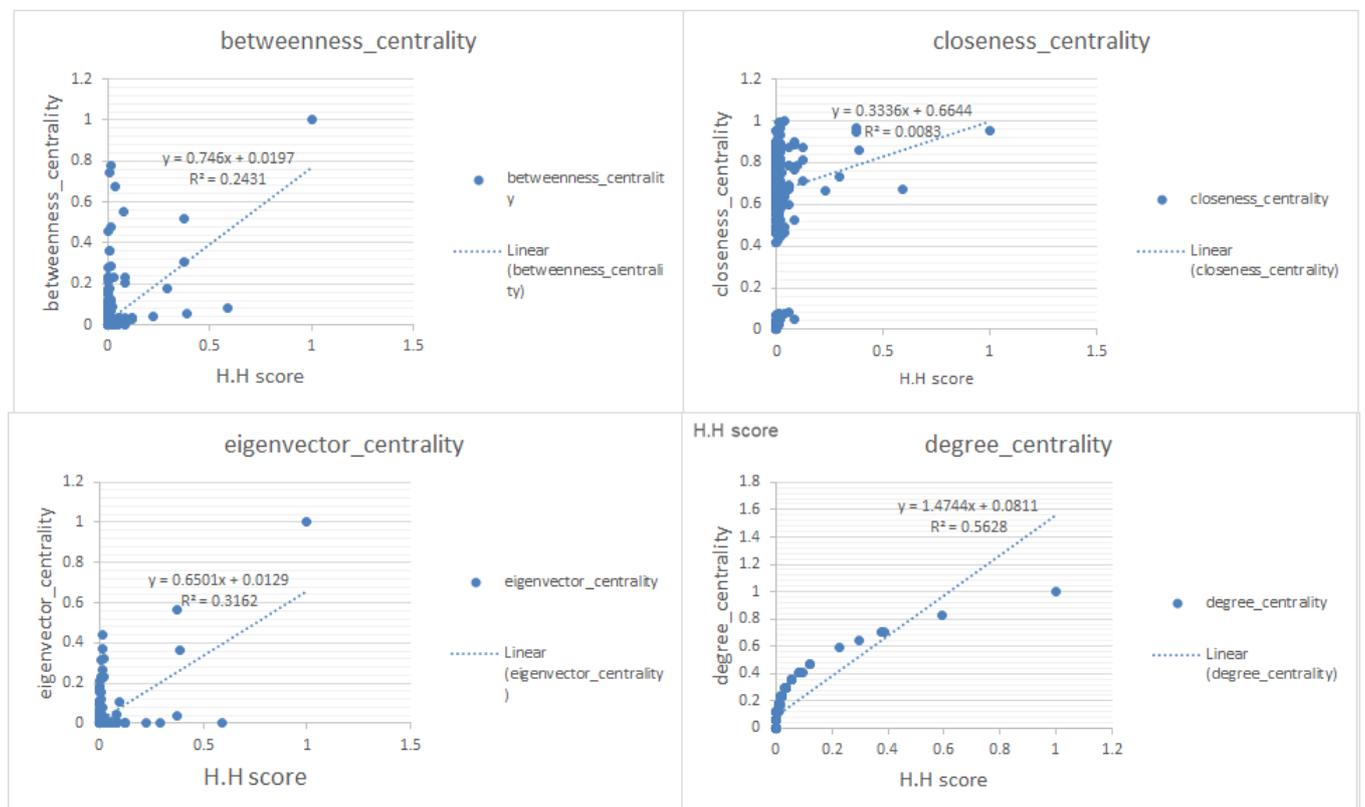

Fig.9. Comparison between H.H and centralities

In this part, we evaluate H.H score and other centralities with respect to changes that will happen in the case of removing nodes with high scores. We first calculate the H.H score for all nodes (in a one-mode projection). Then, from 552 nodes in the crime-crime graph, we select 55 nodes (Top10% of the total number of nodes) with the highest H.H score and remove these nodes from the graph. The number of communities before the removal was 218, which after elimination, reached 159. This means that the number of communities has decreased by 27%. On the other hand, the existing communities have also changed. According to [18], there are six types of network anomalies: born, vanish, grow, merge, split, shrink. The anomalies created by the removal of these nodes are shown in Table 2. 60 communities have vanished, two communities have been shrunk, and one community has split. This means that by eliminating the top 10% of the nodes that have the highest H.H score, 29% of the communities have changed. But, after removing 55 nodes (Top10% of the total number of nodes) that have the lowest H.H score, the communities in the network have not changed. Figure 10 shows community changes caused by removing top10% nodes (55 nodes) in degree centrality, betweenness centrality, closeness centrality, eigenvector centrality, and H.H score.

As you can see, the H.H score has the biggest impact on communities. This measure pays more attention to large communities and gives them a relatively higher score, which is why larger communities disappeared. This has not been addressed by any of the existing centralities, and this shows the importance of the H.H score. The only competitor of H.H score is degree centrality, but the H.H score is better than degree centrality in removing communities of sizes 3 and 4.

Table.2. Abnormalities caused by the removal of nodes with the highest H.H score

| Number of abnormalities | Born | Vanish | Grow | Merge | Split | Shrunk | Rate of communities changes |
|---|---|---|---|---|---|---|---|
| 63 | 0 | 60 | 0 | 0 | 1 | 2 | 29% |

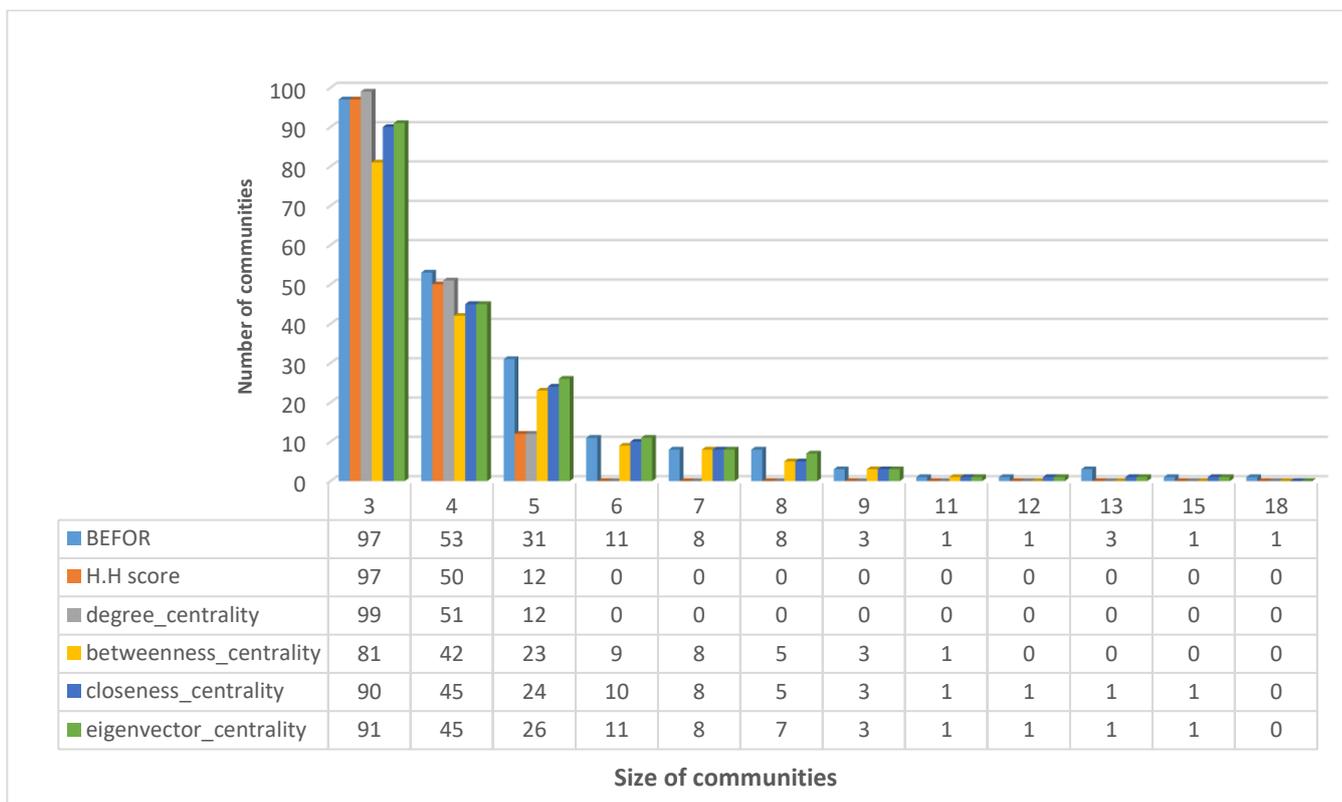

Fig 10. Changes caused by the removal of nodes with a high H.H score

By H.H score, we can find important nodes that have been effective in forming a community, and by removing these nodes, more communities can be eliminated. On the other hand, by adding nodes with a good H.H score, more important and stronger communities can be created in the one-mode projection of a bipartite graph. As shown in Fig.10, this issue has not been addressed by any of the existing centralities. The $R^2$ correlation between centralities and the H.H score is low. For this reason, existing centralities cannot satisfy the three main parameters that are considered by their H.H score. These three are the main parameters for identifying the effective nodes in the formation of communities in one-mode projection of bipartite graphs, and they have sufficient independence to represent the crucial nodes in the construction of the communities.

## IV. FUTURE WORKS

In this paper, we show that by using the H.H score, we can identify the important nodes that have played a significant role in forming communities in a one-mode projection of bipartite graph. The evaluation of the performance of this measure compared to the centralities showed us H.H measure effectively identifies the important nodes in the network (From the perspective of the communities), while the centralities do not have this capability. Also, this score can be extended for un-bipartite, directional and weighted graphs.